# Toward the Optimized Crowdsourcing Strategy for OCR Post-Correction

Omri Suissa[1,2] | Avshalom Elmalech[1] | Maayan Zhitomirsky-Geffet[1]

[1] Bar Ilan University, Department of Information Science, Ramat Gan 52900, Israel

[2] Correspondence email: omrishsu@gmail.com

**Abstract**

Digitization of historical documents is a challenging task in many digital humanities projects. A popular approach for digitization is to scan the documents into images, and then convert images into text using Optical Character Recognition (OCR) algorithms. However, the outcome of OCR processing of historical documents is usually inaccurate and requires post-processing error correction. This study investigates how crowdsourcing can be utilized to correct OCR errors in historical text collections, and which crowdsourcing methodology is the most effective in different scenarios and for various research objectives. A series of experiments with different micro-task's structures and text lengths was conducted with 753 workers on the Amazon's Mechanical Turk platform. The workers had to fix OCR errors in a selected historical text. To analyze the results, new accuracy and efficiency measures have been devised. The analysis suggests that in terms of accuracy, the optimal text length is medium (paragraph-size) and the optimal structure of the experiment is two-phase with a scanned image. In terms of efficiency, the best results were obtained when using longer text in the single-stage structure with no image. The study provides practical recommendations to researchers on how to build the optimal crowdsourcing task for OCR post-correction. The developed methodology can also be utilized to create golden standard historical texts for automatic OCR post-correction. This is the first attempt to systematically investigate the influence of various factors on crowdsourcing-based OCR post-correction and propose an optimal strategy for this process.

## Introduction

In many digital humanities projects, there is a need to automatically analyze the content of large collections of paper-based documents. Digitization of historical text collections is a complex task essential both for research and preservation of cultural heritage. The first step towards this goal is to scan historical documents (books, manuscripts, or newspaper pages) into high-resolution images and then to use an Optical Character Recognition (OCR) technology to convert the images into text. For instance, the Library of Congress has a vast digital historical collection

(https://chroniclingamerica.loc.gov/), which has been digitized using OCR to preserve and make it publicly available. The British Newspaper Archive (https://www.britishnewspaperarchive.co.uk/) maintains an extensive digitized collection with advanced discovery tools (Lansdall-Welfare, Sudhahar, Thompson, Lewis, & Cristianini, 2017). Even commercial enterprises initiated large-scale digitization projects based on OCR technology. For example, the Google Books project has already scanned more than 25 million books and continues to scan 6,000 pages per hour (Heyman, 2015). The quality of the OCR technology is a critical aspect of this process. Unfortunately, OCRed historical texts still contain a significant percentage of errors that undermine further analysis, search, and preservation.

OCR errors come in several forms: insertions, deletions, substitutions, transposition of one or two characters, splitting and concatenation of words, or combination of several error types together in one word (Reynaert, 2008). Many OCR errors are different from human spelling mistakes, and therefore, common spelling correction algorithms are not suitable for the OCR correction task. Moreover, these errors often include punctuation marks, which lead to wrong text segmentation.

In order to automatically correct OCR errors in large historical texts, supervised machine learning algorithms have been applied (Tong & Evans, 1996; Borovikov, Zavorin & Turner, 2004; Bassil & Alwani, 2012; Wick, Ross & Learned-Miller, 2007; Raaijmakers, 2013; Evershed & Fitch, 2014; Kissos & Dershowitz, 2016). Using a corpus of labeled examples (pairs of wrong and right words or sentences) these algorithms employ statistical methods to learn how to distinguish between an error and the correct word or sentence. To create golden standard datasets for training OCR post-correction algorithms for historical texts, crowdsourcing has been employed in previous studies (Von Ahn *et al.*, 2008; Bernstein *et al.*, 2010; Chrons & Sundell, 2011; Volk, Clematide & Furrer, 2016).

In the past decade, numerous studies employed various crowdsourcing approaches with different experimental settings for post-correction of the OCR output (Von Ahn, Maurer, McMillen, Abraham, & Blum, 2008; Chrons & Sundell, 2011; Volk, Clematide, & Furrer, 2016). Yet, it is unclear how to choose the optimal crowdsourcing strategy for this task.

Hence, the goal of this research is to investigate and optimize the crowdsourcing procedure of OCR post-correction in terms of accuracy and efficiency. To this end, a systematic examination of different crowdsourcing approaches was conducted. The considered variables included the task structure, the text length, and supplementary information. The research questions addressed in the study are: 1) How does the structure of the task influence the accuracy and efficiency of crowdsourcing-based OCR post-correction? 2) how does the text length affect the accuracy and efficiency of crowdsourcing-based OCR post-correction? 3) does supplementary information (such as the scanned image of the text) improve the accuracy of crowdsourcing-based OCR post-correction, and how does it affect efficiency? 4) how are the different variables (e.g., text length, task structure, and image presentation) related to the various types of errors in crowdsourcing-based OCR post-correction? 5) based on the above variables, what is the optimal strategy for different scenarios of crowdsourcing-based OCR post-correction?

For the purposes of this study, a series of crowdsourcing experiments with different task structures, text lengths, and supplementary information was carried out with 753 crowd workers recruited using the Amazon's Mechanical Turk (MTurk) platform (https://www.mturk.com/). Overall, 3796 texts of various lengths have been fixed. The developed methodology enables effective creation of golden standard datasets for historical corpora which are required for training machine learning algorithms. To the best of the authors knowledge, this is the first attempt to systematically investigate the influence of various factors on crowdsourcing-based OCR post-correction and propose an optimal strategy for this process. This study has important practical implications for many digital humanities projects which aim to analyze the content of OCRed historical document collections.

## Related Work

### *Crowdsourcing platforms*

Crowdsourcing is based on assigning various well-defined tasks to large groups of non-expert low-paid workers or volunteers. An example of a crowdsourcing platform where workers get paid is Amazon Mechanical Turk (a popular platform with over 100,000 workers), while Zooniverse (https://www.zooniverse.org/) and Crowd4U (https://crowd4u.org) are examples of crowdsourcing platforms which rely on volunteers. It has been shown in previous research that the effect of payment on the outcome quality on crowdsourcing platforms is quite limited (Rogstadius *et al.*, 2011; Aker *et al.*, 2012; Chandler *et al.*, 2012; Gao *et al.*, 2012; Law *et al.*, 2016). It is important to mention that MTurk workers are not dependent on this platform for income (Horton, Rand, & Zeckhauser, 2011; Paolacci, Chandler, & Ipeirotis, 2010). On the other hand, bonuses based on performance had a higher positive effect on the overall work quality. However, these studies also showed that monetary award is not the only incentive of the workers (Mason & Watts, 2010; Aker *et al.*, 2012; Chandler *et al.*, 2012; Gao *et al.*, 2012; Kittur *et al.*, 2013; Yin & Chen, 2015). Psychological factors, like fun and curiosity, influence the work quality as well (Kaufmann & Schulze, 2011; Rogstadius *et al.*, 2011; Law *et al.*, 2016; Elmalech & Grosz, 2017). 69.6% of U.S. based workers consider MTurk as a fruitful way to spend free time (Paolacci, Chandler & Ipeirotis, 2010).

### *Crowdsourcing optimization*

Numerous studies tested optimization strategies for different crowdsourcing tasks, such as article quality assessment (Kittur, Chi & Suh, 2008), extracting keyphrases (Yang *et al.*, 2009), natural language and image annotation (Snow *et al.*, 2008; Sorokin & Forsyth, 2008), facets' extraction (Dakka & Ipeirotis, 2008), and document summarization (El-Haj, Kruschwitz & Fox, 2010). For example, one of the effective optimization techniques is filtering out tasks with low inter-worker agreement (Bernstein et al., 2010; Bigham et al., 2010; Downs, Holbrook, Sheng, & Cranor, 2010; John, Joseph, Horton & Chilton, 2010; Kittur, Smus, Khamkar, & Kraut, 2011). Another popular approach is breaking crowdsourced tasks into micro-tasks or sub-tasks (Bernstein *et al.*, 2010; Kittur *et al.*, 2011). Several studies examined the advantages and disadvantages of crowdsourcing-based research vs. laboratory-based research (Reips, 2000; Kraut *et al.*, 2004; Rand, 2012). They showed that the reliability of the data received from crowdsourcing is similar and sometimes even higher than that of the data collected in laboratory-based experiments (Mason & Suri, 2012; Bartneck *et al.*, 2015). Crowd workers demonstrate a similar level of consistency across multiple personality tests to subjects in the laboratory experiments (Buhrmester, Kwang & Gosling, 2011). In addition, crowd workers are more diverse and thus more representative of the US population than a simple convenience sample of students.

### *Crowdsourcing OCR post-correction*

In the field of OCR post-correction, several studies applied various crowdsourcing approaches with different experimental settings. For instance, Volk, Clematide, & Furrer (2016) presented a scanned image of a document along with the OCRed text and asked crowd workers to correct errors. This approach was found very useful in different languages, but was only tested with a small group of experts. Another common approach is using the post-correction process as a Turing test for websites. In this setting, the website presents only two words every time a user logs in. The first word is known to the website (included in the golden standard dictionary) and the second one is an OCRed word. The user fixes both words, and the benefit is double: the website login process can verify that this is a real human (using the known word) and store an OCR post-correction suggestion (Von Ahn *et al.*, 2008).

When volunteers perform post-correction tasks, a proven method to increase their motivation is a gamification approach (i.e., correcting OCR errors as part of a game) (Chrons & Sundell, 2011). In order to keep the tasks as simple as possible, some studies split the proofing task into three sub-tasks of error finding, fixing, and verifying the corrections (Bernstein *et al.*, 2010). This approach outperforms automatic spellchecking algorithms and significantly improves the quality of the resulting text. Although a considerable amount of research applied crowdsourcing-based methods to OCR post-correction, there is still a need in the systematic and comparative investigation of various factors which influence the quality and efficiency of this approach.

# Research Method

*The corpus of the study*

In the conducted experiments, a digital version of “The Joyful Heart” by Robert Haven Schauffler (written in 1914) was used as a representative historical text corpus. The reason for choosing the Joyful Heart text is twofold. First, this corpus contains some vocabulary which is uncommon in modern English, while it still can be understood by crowd workers. Therefore, the task is more challenging and requires the full worker’s attention. Second, crowd workers are usually not familiar with this text, and thus, there is no effect or bias related to previous experience with the proofread content.

Several types of OCR errors have been examined in the literature (Fernandes, Santos & Milidiú, 2010):

1. **Real word errors** are words that are spelled correctly (from a linguistics point of view), but do not match the scanned image. For example, by replacing the character **c** with the character **b** in the word **cat,** the word **bat** is formed**,** which is a real English word.
2. **Non-real word errors** are “words” that are misspelled. For example, by replacing the character **c** with the character **a** in the word **cat,** the word **aat** is obtained, which does not exist in English.
3. **Non-word errors** are “words” that contain digits and other non-alphabetic characters. For example, replacing the character **t** with the character **4** in the word **cat,** results in **ca4**, which is not a word.
4. **New line errors** occur when a word is split into two separate words by a new line (due to the width limit of the page), such as "c-at".
5. **Tokenization errors** occur when a single word is split into multiple words with a white space or multiple words are mistakenly concatenated into a single word, for example, "ac at" instead of "a cat".

Next, based on the previous work (Reynaert, 2008), typical OCR errors were artificially generated and inserted to the original text to simulate OCRed text, using the following algorithm:

**Input:** The original correct gold standard text (GSt), Noise ratio (NR), $0 < NR < 1$.

**Output:** The artificially OCRed text (OCRt)

```
for each line in GSt do
        r1 = random number between 0 to 1
        if r1 < NR then
                Remove one random character from the line
```

```
        end if
        r2 = random number between 0 to 1
        if r2 < NR then
                Insert one random character to one random position in the line
        end if
        r3 = random number between 0 to 1
        if r3 < NR then
                Substitute two random characters in one random position in the line
                (randomly repeat once or twice)
        end if
        OCRt <= OCRt + line
end for each
return OCRt
```

Then, the text was split into sentences, paragraphs, and articles. Each paragraph was constructed from twenty consecutive sentences (on average), and each article comprised three (not necessarily consecutive) paragraphs. Finally, images were created for each article using an old newspaper background to visually imitate scanned images.

## *Experimental Setting*

### *The study population*

A total number of 753 crowd workers were recruited on Amazon Mechanical Turk. Table 1 presents the crowd workers' demographic characteristics. Most of the crowd workers were 21 to 40 years old. 73% of them were US residents similarly to previous findings (Difallah, Filatova and Ipeirotis, 2018), and 20% of the crowd workers were located in India. 50% of them self-identified as male, 47% as female and 3% as others.

All participants signed a letter of consent before participating in the experiment. The letter included a general description of the research procedure and emphasized that all the collected data is anonymized and stored in a secure encrypted database. Participants were able to quit at any given point of the experiment. This study has been approved by the IRB of the Faculty of Humanities at Bar-Ilan University.

| **Variable** | | Count | N(%) |
|---|---|---|---|
| **Age** | 0-20 | 132 | 17.5 |
| | 21-30 | 213 | 28.2 |
| | 31-40 | 176 | 23.3 |
| | 41-50 | 121 | 16.0 |
| | 51+ | 111 | 14.7 |
| **Nationality** | U.S | 548 | 72.7 |
| | India | 148 | 19.6 |
| | Other | 57 | 7.5 |
| **Gender** | Male | 375 | 49.8 |
| | Female | 355 | 47.1 |
| | Other | 23 | 3.0 |

**Table 1:** Crowd-workers' demographic characteristics.

*Structures of Crowdsourcing Micro-tasks*

This study compared two types of task structure: 1) a two-phase "Find-Fix" procedure based on (Bernstein *et al.*, 2010), and 2) a single-phase "Proofing" procedure. Accordingly, one third of the crowd workers was asked to find errors in the text (the "Find-Only" group), another third of the workers was asked to fix the errors found by the first group (the "Fix-Only" group), and for comparison, the last group was asked to detect and immediately fix errors in the same session (the "Proofing" group). Each of the crowd workers in every group was randomly assigned one of the six error correction tasks with three different text lengths, as follows: 1) three articles, or 2) three paragraphs, or 3) twenty sentences on average, which were either displayed with the corresponding “scanned” image of the original article or without it. All automatic spelling error detection and correction functions were disabled during the experiments. Each worker was paid 60 cents for completing a single task. Figure 1-3 illustrate the micro-task structures of the experiments for the "Find-Only", "Fix-Only", and "Proofing" groups, respectively.

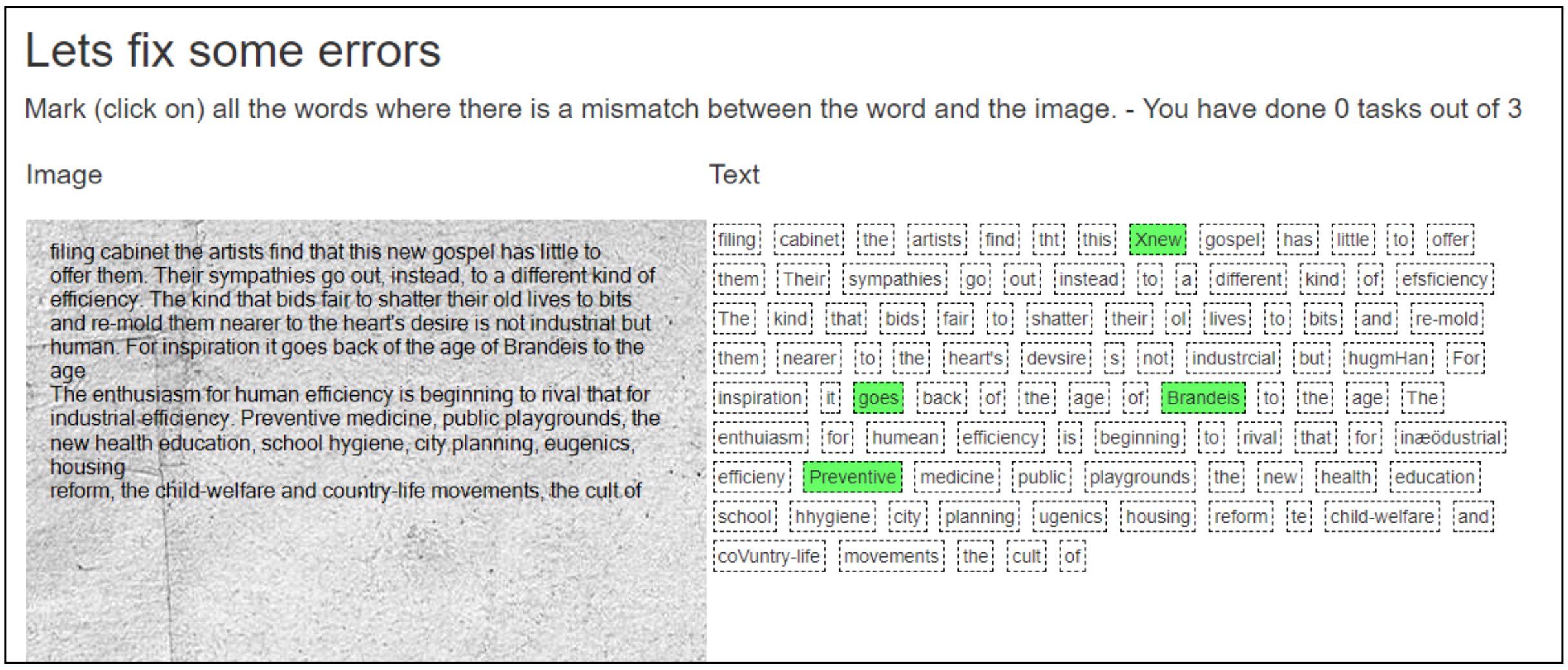

**Figure 1**: The experiment's structure ("Find-Only" group) presenting the “scanned” image and the text of a paragraph.

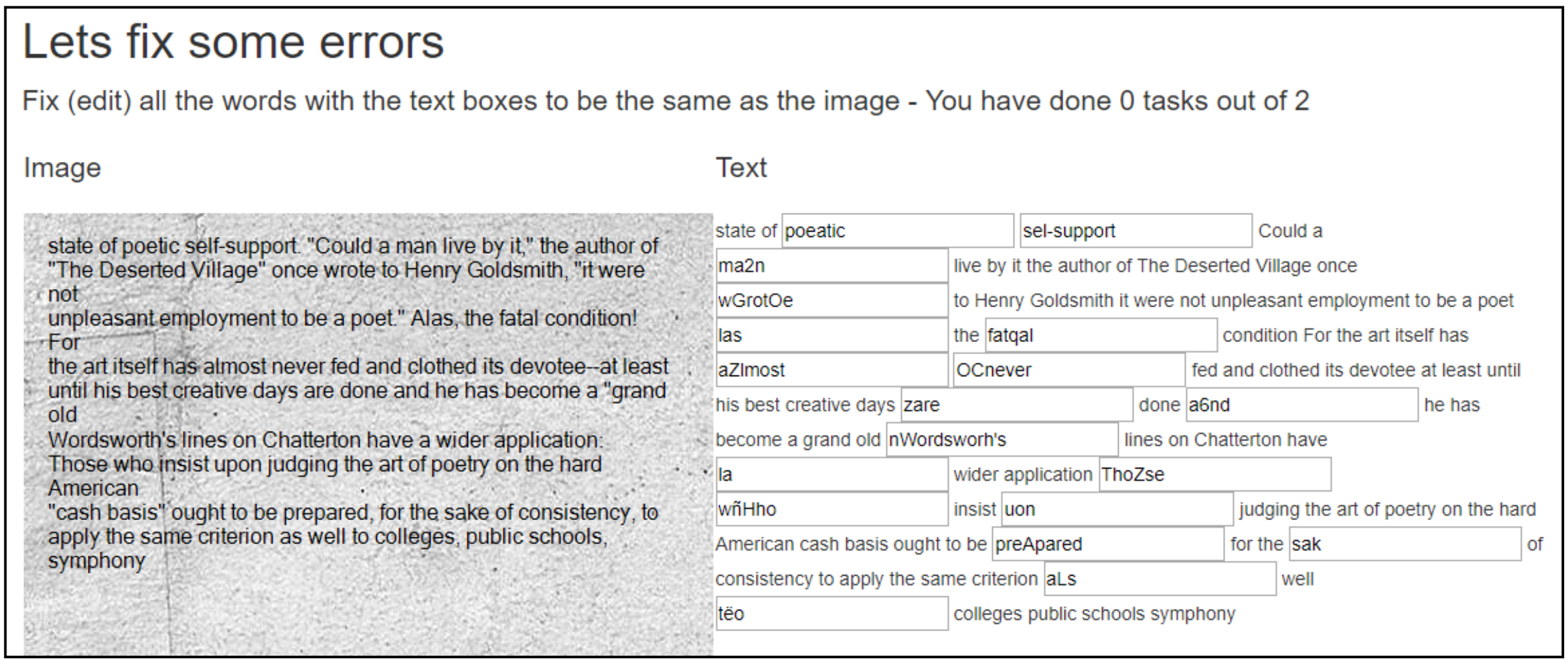

**Figure 2**: The experiment's structure ("Fix-Only" group) presenting the “scanned” image and the text of a paragraph.

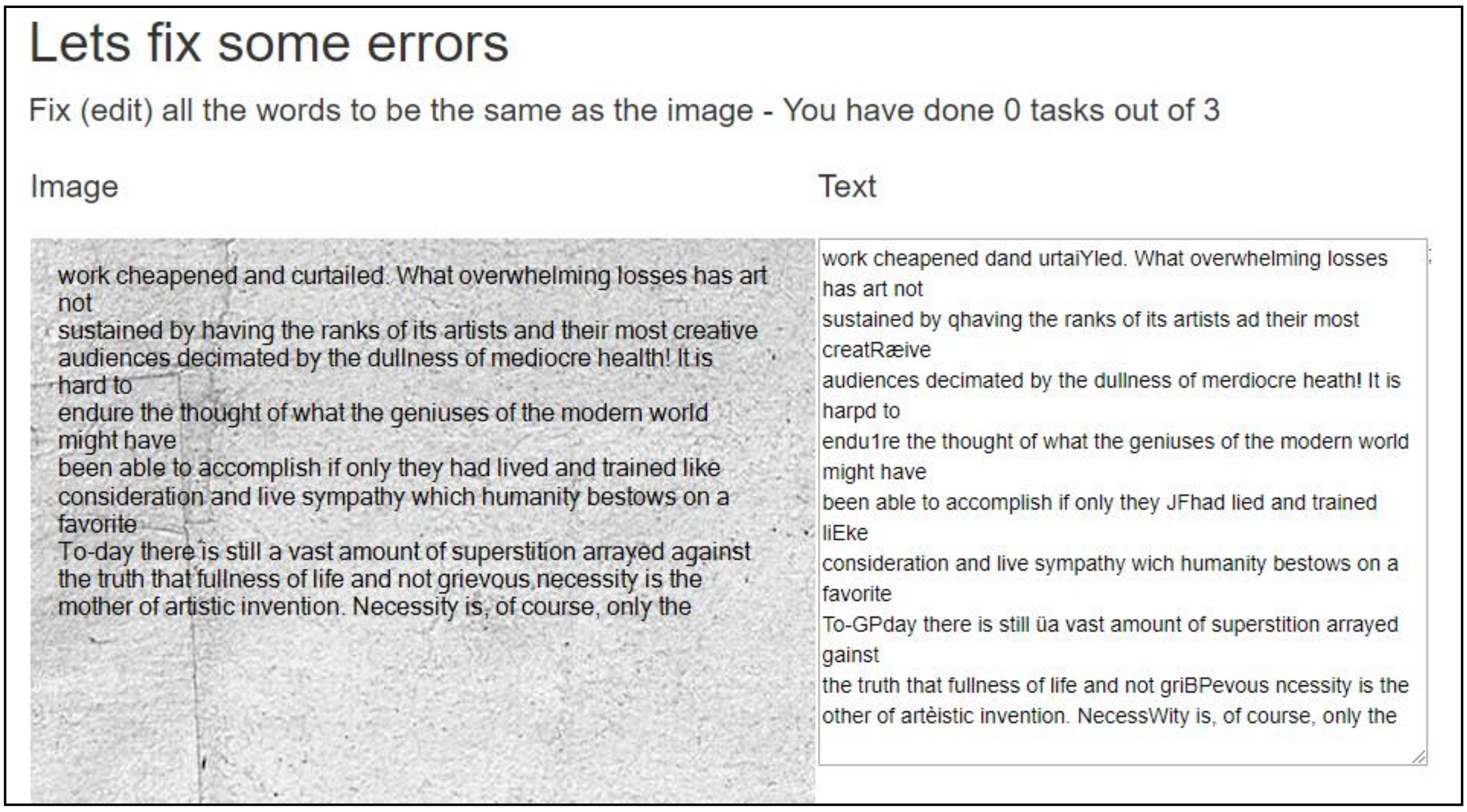

**Figure 3**: The experiment's structure ("Proofing" group) presenting the “scanned” image and the text of a paragraph.

## *Evaluation Metrics*

To assess the quality of the results, both workers’ efficiency in completing the tasks and accuracy of their corrections were measured.

### *Accuracy evaluation*

Basically, the task's accuracy is the ratio between the number of worker’s corrections and the initial number of errors in the OCRed text. The number of worker’s corrections is calculated as a difference between the minimal edit distance (Levenshtein, 1966), denoted as *mindist*, of the *OCRed* text from the (correct) golden standard version of the text, *GS*, and the minimal edit distance of the fixed text, *Fixed*, (after the workers’ corrections) from the golden standard text. The initial number of errors in the OCRed text is computed as the minimal edit distance between the OCRed text and the golden standard text. Accuracy is calculated only for the "Fix-Only" and "Proofing" groups, since "Fix-Only" group includes the contribution of the "Find-Only" group.

More formally, the task's accuracy is calculated as follows:

$$(Eq\ 3)\ acc_{task} = \begin{cases} \dfrac{mindist_{GS,OCRed} - mindist_{GS,Fixed}}{mindist_{GS,OCRed}}, & mindist_{GS,OCRed} \geq mindist_{GS,Fixed} \\ 0, & otherwise. \end{cases}$$

If $mindist_{GS,OCRed}$ is smaller than $mindist_{GS,Fixed}$ (i.e. a crowd worker actually made the text even worse than the given OCRed version), the metric value is set to 0. Therefore, the output value is the improvement ratio of the crowd worker's accuracy.

Basically, $mindist_{t1,t2}$ is the sum of the minimal number of substituted characters, *S*, the minimal number of deleted characters, *D*, and the minimal number of inserted characters, *I*, required to transform the output text into the target text.

More formally, $mindist_{t1,t2}$ is defined as follows:

$(Eq\ 4)\ mindist_{t1,t2} = \frac{I+S+D}{N}$, where *N* is the number of characters in *t1*.

*mindist* reflects the nature of the OCR processing, which is based on individual character identification.

*Word-based Error Analysis*

To complete the character-based accuracy evaluation, an error analysis based on words was performed. The word-based analysis is important from the user perspective, since users comprehend texts and search them by whole words.

*Proofing structure errors*

In the Proofing approach, there are two potential error types that a crowd worker can make:

1) Miss: missing (not fixing) an OCR error (a spelling mistake).
2) Wrong: wrong correction of a word (inserting a mistake in the correct word).

The first step in calculating a word-based metric is to identify the words in the OCRed text. The space character, a common word delimiter, is unreliable in OCRed documents as it is usually missed by OCR algorithms causing multiple word concatenation (e.g. "forexample") and sometimes whole words are mistakenly split into parts by wrongly identified and inserted space characters (e.g., "exam ple"). Therefore, to estimate word-based error rates for the Proofing method, a global alignment algorithm for parallel texts was employed. To this end, Needleman–Wunsch's alignment algorithm (Needle & Christus, 1970) was implemented, which was initially invented to align proteins and nucleotides sequences in bioinformatics and was later found to be effective for NLP tasks. To calculate the number of Proofing-Miss errors, the alignment algorithm was run twice. First, for the OCRed and GS texts, to detect incorrect words in the OCRed text (with no alignment to GS), denoted as *OCRE*, which are supposed to be corrected by the crowd workers. Second, for the Fixed and GS text, to find incorrect words in the Fixed text (with no alignment in GS), denoted as *FE*. Then, the missed corrections (i.e., the number of Proofing-Miss errors) are calculated as the size of the intersection of *OCRE* and *FE*. The result is divided by the overall number of errors of both types (Proofing-Miss and Proofing-Wrong), *Errors*, to compute the relative part of the Proofing-Miss errors out of the total amount of errors.

$(Eq\ 5) Proofing - Miss = \frac{|OCRE \cap FE|}{Errors}$ , where *FE* is the number of word errors in the Fixed text, *OCRE* is the number of word errors in the OCRed text, *Errors* is the number of all the errors of all types (sum of Eq5, Eq6).

The relative amount of the Proofing-Wrong errors was computed as the number of words in *FE* that do not appear in *OCRE*, divided by *Errors*. In other words, these are words that were correct in the OCRed text, but were mistakenly modified by the workers.

$(Eq\ 6) Proofing - Wrong = \frac{|FE| - |OCRE \cap FE|}{Errors}$, where *FE* is the number of word errors in the Fixed text and *OCRE* is the number of word errors in the OCRed text. *Errors* is the number of all the errors of all types (sum of Eq5, Eq6).

*Find-Fix Structure Errors*

In the Find-Fix structure, there are three potential error types that a crowd worker can make:

1) Find-Miss: in the Find stage, missing (not selecting) an OCR error.
2) Find-Wrong: in the Find stage, selecting a correct word as an error.
3) Fix-Wrong: in the Fix stage, a wrong correction of a word that was marked in the Find stage as an error.

At the first phase of the Find-Fix experiment, a worker selects words with OCR errors, and in the second phase, another worker corrects these words. Since workers identify and mark the entire words (rather than individual characters), this experimental structure simplifies the word-based error calculation. To this end, a set of words selected as errors by a worker was denoted as *WSE*, then the number of wrongly selected words (*Find-Wrong* errors) was computed as the size of the intersection of *WSE* and *GS* (a set of words in the golden standard text). The intersection is assessed by a simple search of the selected word (as a separate word) in GS. Correctly selected words are the rest of the words in *WSE*. Note, that selected adjacent words were counted as a single word if they were found in GS after concatenation. A set of the edited words at the Fix stage was denoted as *EW*. Then, correct edits (*CR*) are words in the intersection of *EW* and *GS*. Wrong edits are therefore the rest of the words in *EW*. It should be noticed that the above error estimation metrics might produce slightly inaccurate results, when wrongly edited words appear somewhere in the GS text. Finally, the number of the words with errors which were missed at the Find stage is computed as the difference between the size of *OCRE* and the number of the correctly selected words. *OCRE* (estimated as for Proofing error analysis above) is the number of word errors in the OCRed text. More formally, the error rates are measured as follows:

$(Eq\ 7) Find - Wrong = \frac{|WSE \cap GS|}{Errors}$, where *WSE* (worker's selections) is the number of words selected by a worker at the Find phase, and *Errors* is the overall number of errors of all types for the Find-Fix experiment (sum of Eq7, Eq8, Eq9).

$(Eq\ 8) Fix - Wrong = \frac{|EW| - |EW \cap GS|}{Errors}$, where *EW* (worker's edits) is the number of edits of (previously selected) words at the Fix phase of the experiment and *Errors* is the overall number of errors of all types for the Find-Fix experiment (sum of Eq7, Eq8, Eq9).

$(Eq\ 9) Find - Miss = \frac{|OCRE| - (|WSE| - |WSE \cap GS|)}{Errors}$, where *WSE* (worker's selections) is the number of selections of words with errors, *OCRE* is the number of word errors in OCRed text and *Errors* is the overall number of errors of all types for the Find-Fix experiment (sum of Eq7, Eq8, Eq9).

### *Efficiency evaluation*

The efficiency evaluation metric is the period (in seconds) it took a crowd worker to complete the task, divided by the length (the number of characters) of the task's text. For the two-phase "Find-Fix" strategy, the task efficiency for each phase in separate was calculated, and then the results were summed. More formally, for a task *t*, task type *(*where *fix* stands for the fix-only group*, find* is for the find-only group, *p* is for the proofing group*)* and the length of the original text (gold standard text length - $GS_{len}$), the work efficiency is calculated as follows:

$$(Eq\ 1)\ time_{task_{fix+find}} = \frac{\left(time_{end_{fix}} - time_{start_{fix}}\right) + \left(time_{end_{find}} - time_{start_{find}}\right)}{GS_{len}}$$

$$(Eq\ 2)\ time_{task_p} = \frac{\left(time_{end_p} - time_{start_p}\right)}{GS_{len}}$$

These measures allow to compare the strategies and analyze the effect of the text length, task structure, and displaying the image on the performance of the crowd worker. Since crowd workers are often paid according to the time needed to perform a task, efficiency evaluation also reflects the cost of the work.

## Results

*The accuracy of corrections*

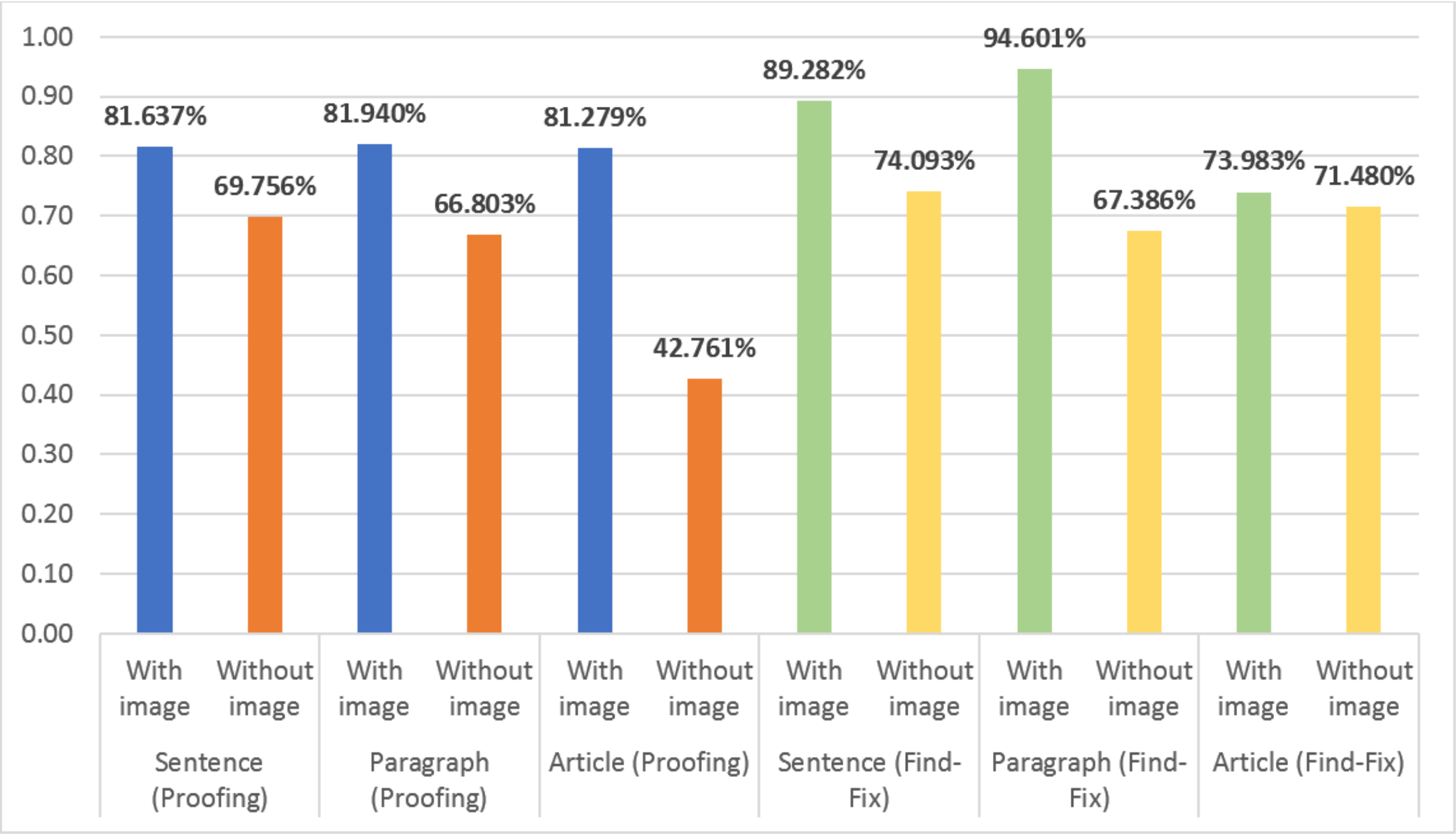

**Figure 4:** The average accuracy values for various types of task's structures (the higher the values, the more accurate the workers' corrections).

Figure 4 displays the accuracy values at the character level for the various task's structures. As can be observed from the figure, Find-Fix outperformed Proofing both with and without the scanned image (using one-way Anova test: $F_{(1,1931)}$=33.79, $p<0.001$ and $F_{(1,1861)}$=8.126, $p<0.05$, respectively). This result shows that workers perform better in simpler tasks. Presenting an image improved the accuracy of the work in both Proofing ($F_{(1,1931)}$=33.789, $p<0.001$) and Find-Fix approaches ($F_{(1,1892)}$=123.52, $p<0.001$).

The text's length had a significant impact on accuracy for Proofing without an image ($F_{(2,1899)}$=5.178, $p<0.01$). In this experimental setting, accuracy was much higher for shorter texts (69.76% and 66.8%), while for full articles, this approach yielded the lowest accuracy value (42.61%). Possibly, workers tend to lose attention after the first paragraph and thus make more errors for longer texts. These findings also reflect the relation between the length of the text and the need for using an image. For proofing of longer texts, displaying the image significantly improves the accuracy of the results. Thus, when images were presented, proofing of full articles was virtually as accurate as of sentences and paragraphs.

For Find-Fix with an image, the text's length also played a significant role ($F_{(2,1923)}$=13.055, $p<0.001$), with the best result of the study produced for paragraphs (94.6%). A reason for this finding could be that while providing some context improves worker's accuracy, too much context (like in a full article) causes a loss of attention and increases the number of errors. The lowest accuracy value for Find-Fix with images was obtained for full articles (73.98%), which was also the only case with no significant difference between the Proofing and Find-Fix approaches. A possible explanation for these results might be the workers' tendency to use the image only in more complex tasks (Proofing a complete article), while with simpler tasks (Find or Fix in separate) workers prefer to save time and effort required for examining the image. This may contradict Bernstein's et al. (2010) methodology and findings for this typical scenario. Bernstein et al. (2010) suggest that Find-Fix-Verify is a better crowdsourcing task design than Proofing for summarization and spelling error correction. The experiment in this study confirms this result in most cases, but in the case of long texts (articles of 3 paragraphs) with an OCR image, it has been found that the naïve proofing method achieves better results. Note that the current study's experiment does not include the Verify step suggested by Bernstein et al. (2010).

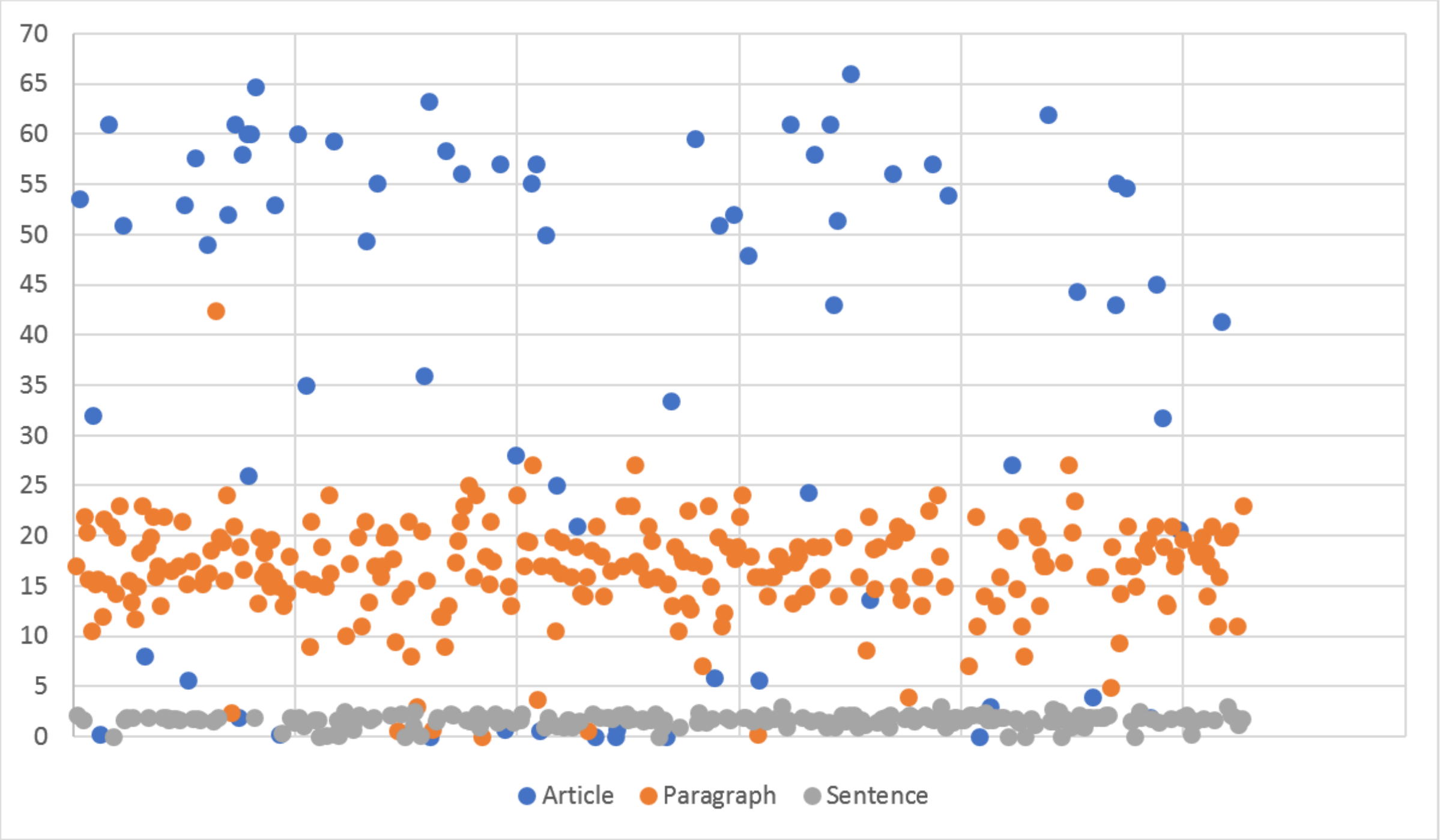

**Figure 5:** OCR corrections distribution per worker and text length.

To analyze the individual worker's performance, Figure 5 presents the distribution of the average number of corrections of each crowd worker for various text lengths. As expected for the majority of workers there were similar correction rates (1-5 for sentences, 10-25 for paragraphs and 50-65 for full articles), while for longer texts there was a greater variance of error corrections. Next, the different error types that occurred in the study's experiments were analyzed.

### *Word-based Error Analysis*

The average length of an article was 314 words. Each paragraph contained 105 words on average, and each sentence contained 11 words on average. Error analysis for the different types of experiments and errors was conducted, and the following results were obtained.

#### *Find-Fix Structure Errors*

As mentioned above, there are three potential error types that a crowd worker can make in the Find-Fix scenario: Find-Miss, Find-Wrong, and Fix-Wrong.

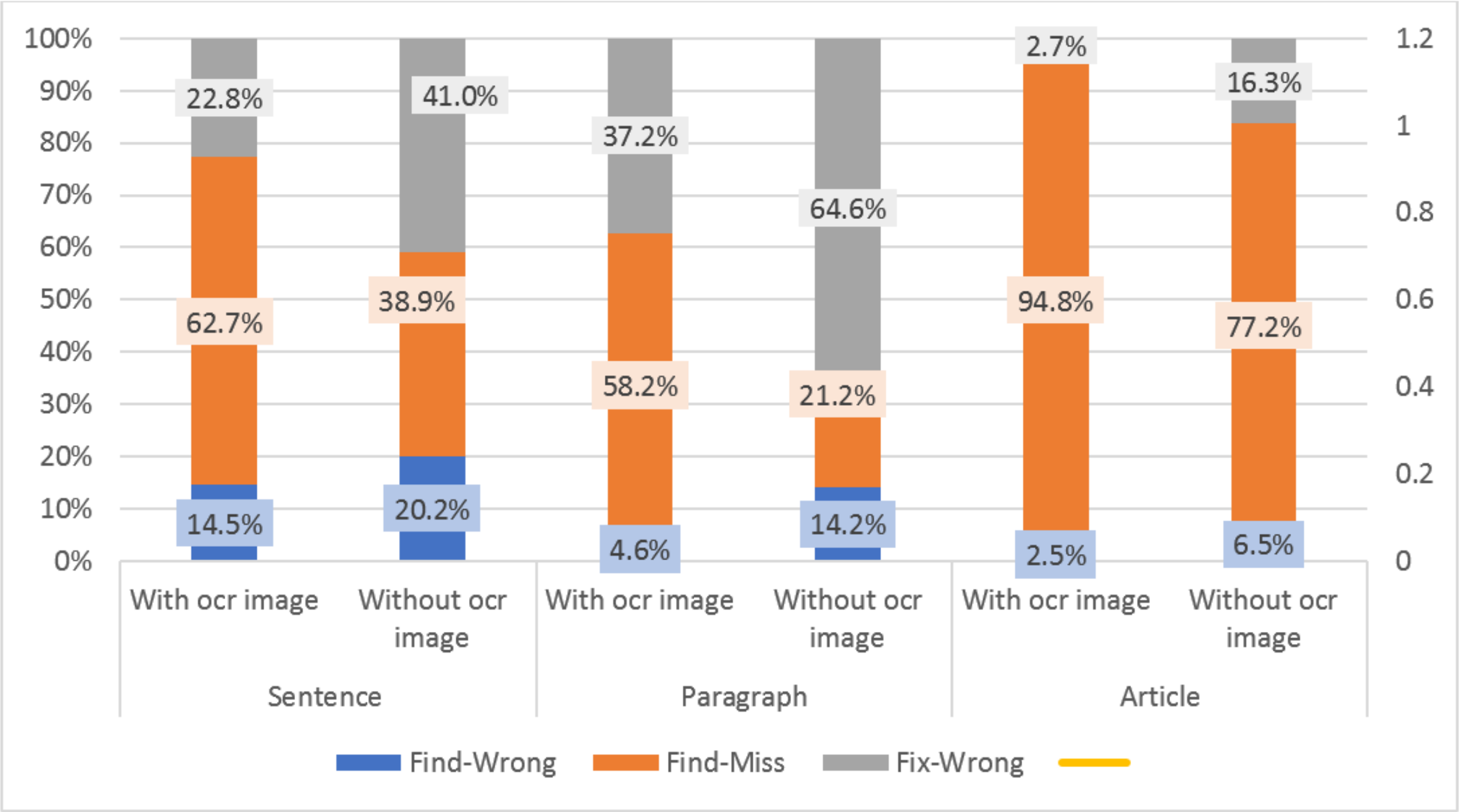

**Figure 6:** The distribution of errors using the Find-Fix method at the word level.

As can be observed in Figure 6, when presenting the scanned image the majority of errors are of the Find-Miss type for all text lengths, while workers tend to miss relatively more OCR errors (Find-Miss) when fixing sentences and articles than paragraphs ($F_{(2,979)}$=19.182, p<0.001). Thus, the lowest percentage of Miss type errors was obtained for paragraphs, which also showed the best accuracy results as described above. Interestingly, presenting the image reduces the percentage of wrong identification of correct words as OCR errors (Find-Wrong) and spelling errors while fixing the text (Fix-Wrong) for paragraphs ($F_{(2,389)}$=42.323, p<0.001; $F_{(2,384)}$=120.007, p<0.001, respectively) and articles ($F_{(2,110)}$=5.842, p<0.05;$F_{(2,108)}$=27.672, p<0.001, respectively). The rarest error type is Find-Wrong (identifying a correct word as an error) for all the experimental structures. Moreover, articles contain a higher percentage of Find-Miss mistakes, which suggests that crowd workers lose attention when correcting long texts.

*Proofing Structure Errors*

In the Proofing approach, there are two potential error types that a crowd worker can make: Proofing-Miss and Proofing-Wrong.

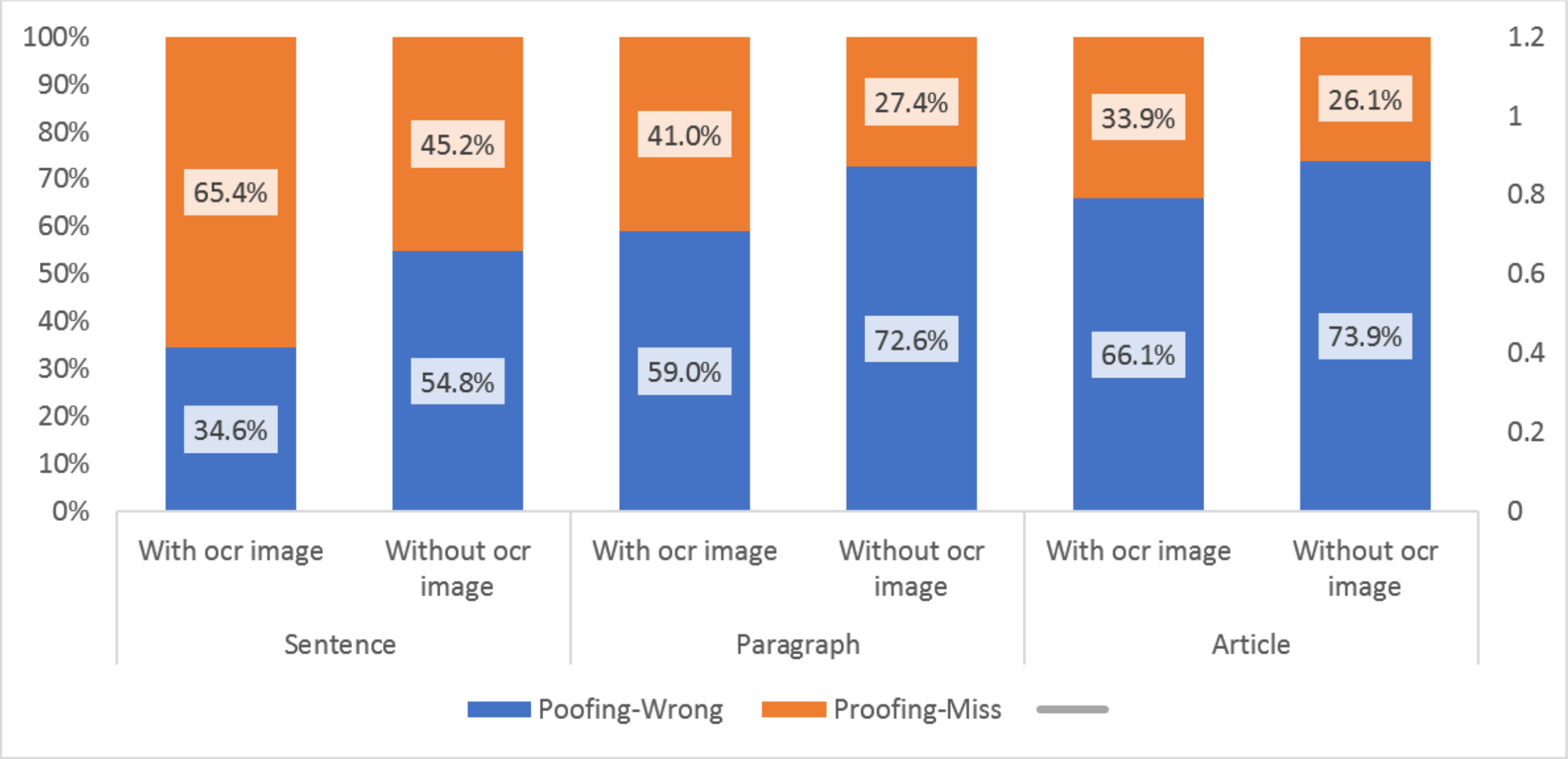

**Figure 7:** The distribution of various types of mistakes using the Proofing structure.

As can be observed from Figure 7, the longer the text, the more wrong corrections were produced by the workers. As opposed to the Find-Fix structure, the majority of errors in Proofing was wrong corrections (except for the case of sentences with images). This leads to the conclusion that the Find-Fix approach helps reduce the percentage of wrong corrections compared to the Proofing approach. In addition, for Proofing, as for the Find-Fix structure, presenting the image reduces the total number of wrong corrections ($F=29.999$, $p<0.001$) and increases the percentage of Miss type errors for all text lengths.

*The crowd work efficiency*

Figure 8 displays the average efficiency for the various task's structures. As can be observed from the figure, Proofing approach outperformed the Find-Fix approach ($F_{(1,3868)}=78.48$, $p<0.001$). The differences between the approaches could be explained by the overhead of the Find-Fix structure, which requires to read the text twice (once at the Find stage and then by another worker at the Fix stage). However, the efficiency of the Find-Fix method significantly increases for longer texts ($F_{(2,1961)}=1194$, $p<0.001$). This can be explained by the assumption that the Fix stage does not require reading the entire text again, but only viewing the close context of the marked errors. Therefore, for full articles, there is no significant difference among the two approaches since the number of the marked errors (and their close contexts) is very small compared to the overall length of an article.

Presenting the scanned image along with the text, had a significant negative effect on the time spent on the task ($F_{(1,3864)}$=6.81, p<0.01). However, presenting the image had no significant effect on the task efficiency when fixing sentences, which shows that the cost of using the image in short texts is low and does not affect the overall time of the task.

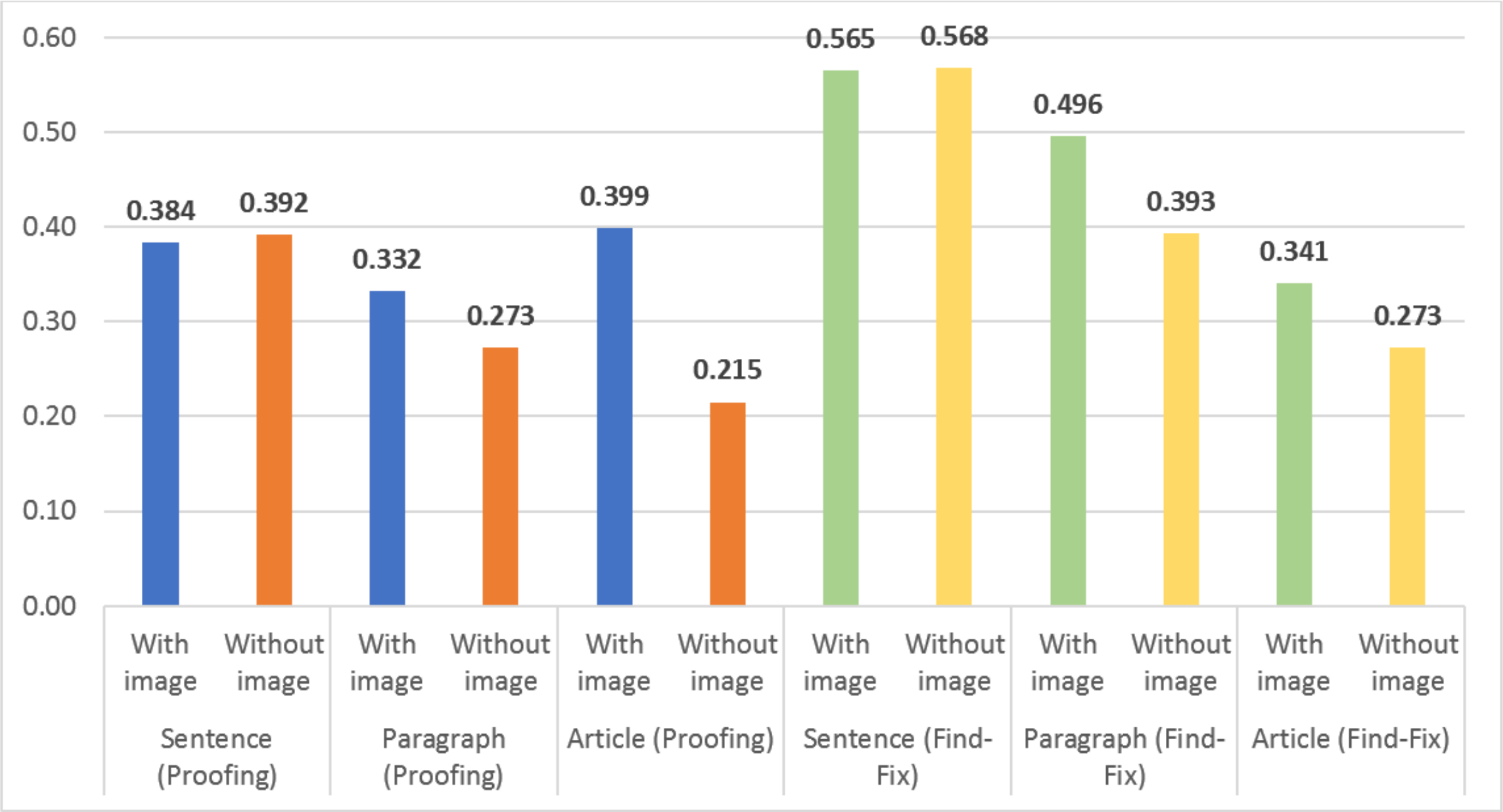

**Figure 8:** The average efficiency values for various types of tasks (the lower the values, the higher the efficiency).

## Discussion and Conclusions

This study conducted a systematic analysis of various crowdsourcing methodologies in order to investigate the effect of different variables on the accuracy and efficiency of post-correction of OCR errors. New evaluation measures were devised to compare the different task structures and propose the optimal strategy for various scenarios and research goals.

The obtained results suggest that there are significant costs and benefits when presenting the scanned image. On the one hand, presenting the image helps reduce the Miss type errors and increase the overall accuracy of correction. On the other hand, the efficiency declines when workers use images, although they seem to ignore the image when the cost (effort) is high, and the benefit is low (when completing simple tasks on long texts). Moreover, crowd workers seem to lose attention when proofing long texts, especially when they cannot view the image.

Regarding the text's length, it seems that paragraphs provide the optimal contextual balance since they contain enough semantic context to allow the worker to achieve high-quality corrections without losing attention. In addition, the two-stage Find-Fix structure is, in most cases, the optimal approach in terms of accuracy, while its disadvantage in terms of efficiency is neutralized for long texts. Overall, for long texts (full articles) accuracy is lower than for shorter texts, but the differences in performance (for both accuracy and efficiency) of the two approaches (Find-Fix vs. Proofing) are minimal.

Based on these findings, researchers can optimize the process to achieve better results according to their objectives and priorities. The recommendations depend on the goals of the crowdsourcing campaign, such as: 1) improving accuracy; 2) improving efficiency; 3) reducing the percentage of each type of errors. Figure 9 shows the optimal strategy decision tree for maximizing the accuracy of the results.

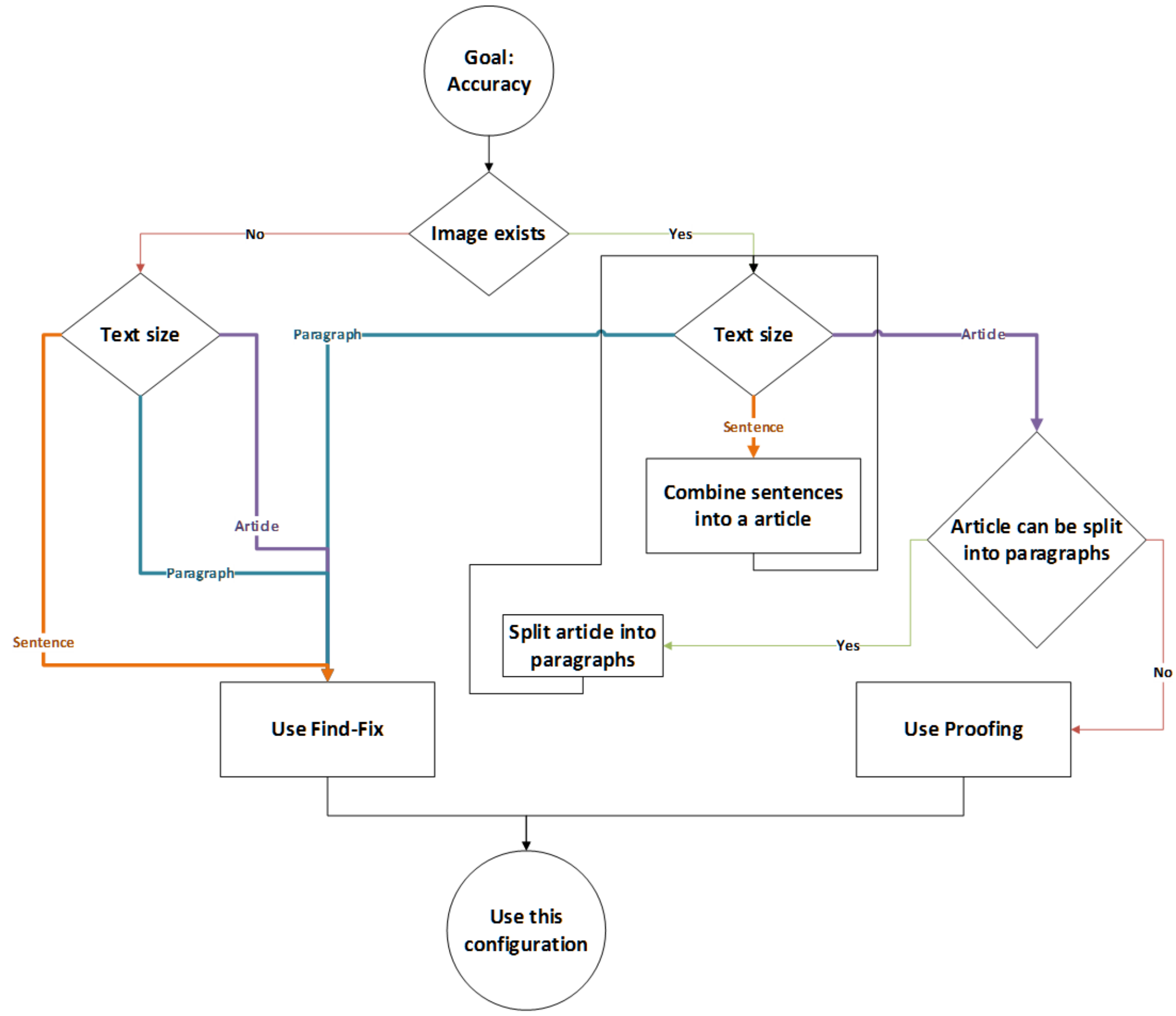

**Figure 9:** The optimal strategy decision tree for accuracy.

This strategy is suitable when there is no particular limit on time and budget, which is typical for small-scale projects. As shown in the diagram, the first step of the strategy selection process depends on the existence of the scanned image of the text. If the image exists, it should be displayed to crowd workers to improve the accuracy of corrections. The second step is to examine the length of the text. As the optimal length of the supplied context is a paragraph, long articles should be split into paragraphs. Note that accurate automatic segmentation of OCRed texts might be hard to achieve since punctuation marks are often ignored or mistakenly identified as letters by OCR algorithms. Therefore, the case when an article could not be split into paragraphs has been considered. If the image cannot be provided, there is no need to split the article, as no significant influence of length on the accuracy of corrections was found in this case. The two-phase (Find-Fix) task structure should be used in all scenarios, except for full articles that cannot be split into paragraphs, where Proofing was found more effective.

When using the Find-Fix structure, an error minimization strategy can be used, as shown in Figure 10. The strategy depends on the type of error to be minimized and on the text length. In addition, the image is used only for Find-Wrong and Fix-Wrong errors' reduction.

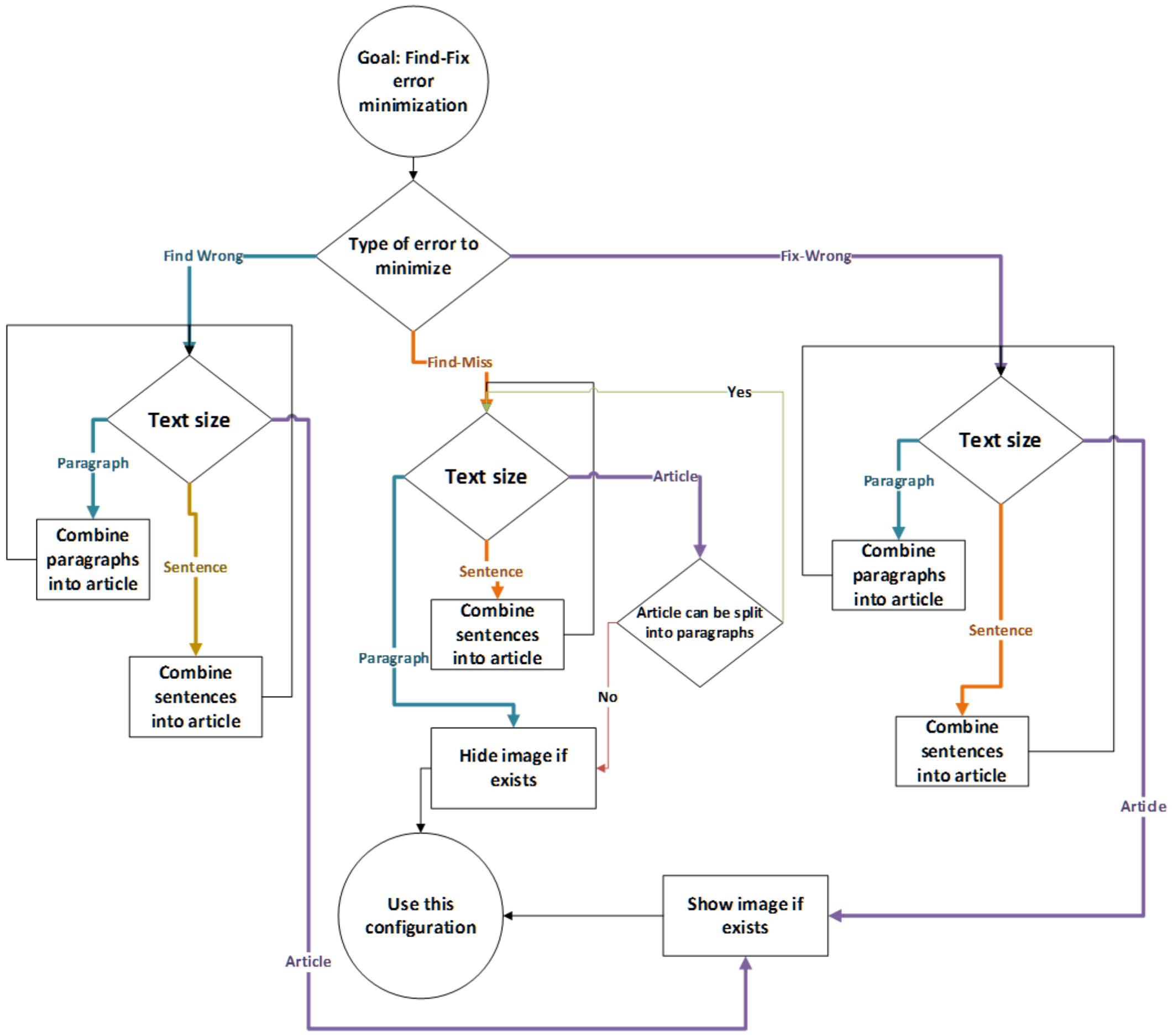

**Figure 10:** The error minimization strategy decision tree for accuracy using find-fix methodology.

When using the Proofing structure, an error minimization strategy can be used, as shown in Figure 11. Similarly, the strategy depends on the error type and text length, while the image should be displayed only for reducing Proofing-Wrong errors.

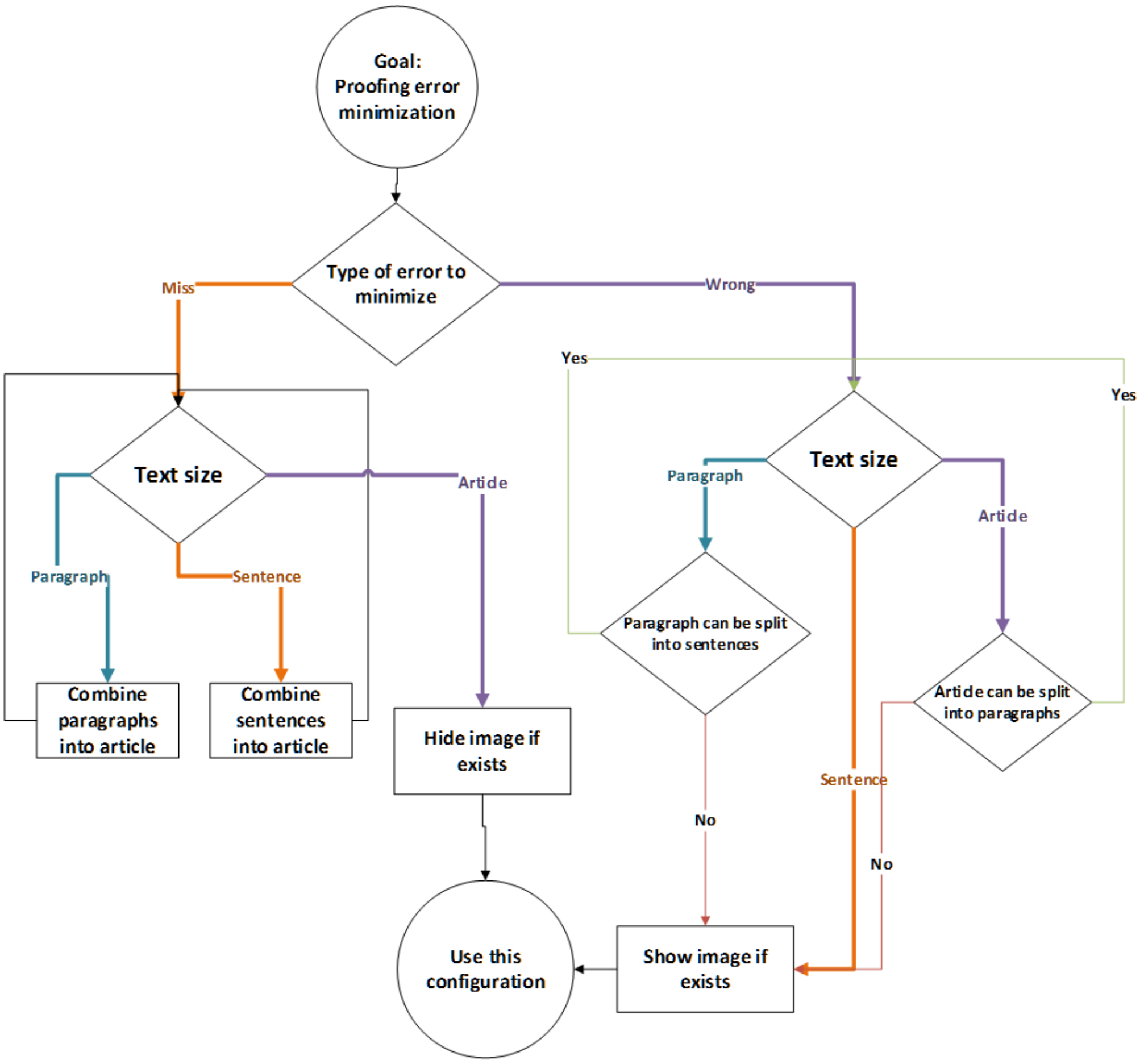

**Figure 11:** The error minimization strategy decision tree for accuracy using proofing methodology.

Figure 12 shows the optimal strategy decision tree when the goal is to increase the efficiency of the task.

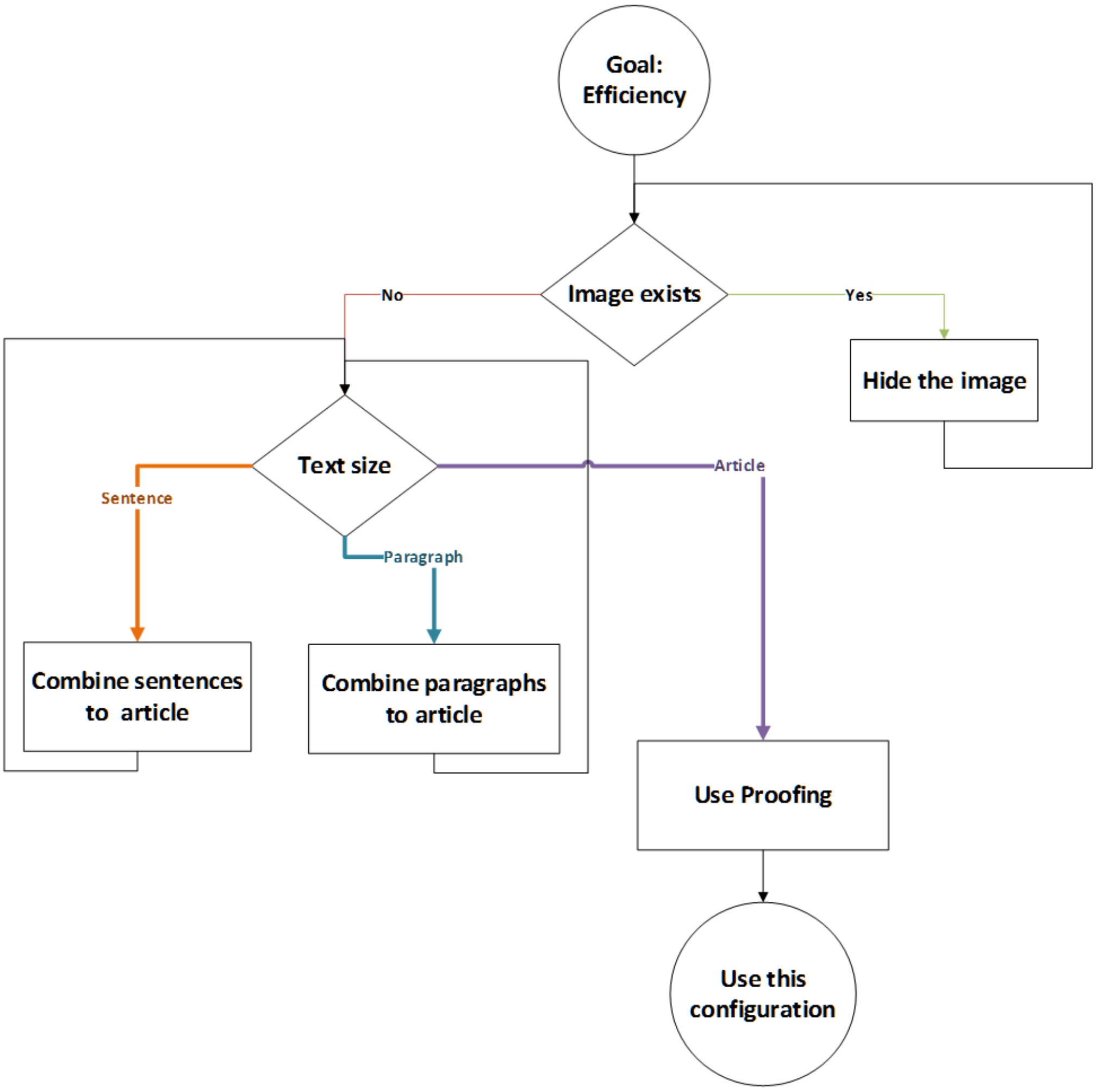

**Figure 12:** The optimal strategy decision tree for efficiency.

It is important to note that in many crowdsourcing tasks, saving time can be directly translated into reducing the cost. This strategy is suitable when time and budget are the most important factors, which is typical for large-scale projects. The proposed strategy suggests that hiding the image even if it is available and using a single-stage Proofing structure of the task is beneficial in terms of efficiency, as a two-stage structure requires additional (partial) reads of the text.

The corrected corpus of OCRed texts created by the optimized crowdsourcing procedure can serve as a training dataset for machine learning algorithms. This study helps reduce the complexity of the crowdsourcing strategy choice and, hence, has important practical implications for many digital humanities projects which aim to digitize and analyze the content of OCRed document collections.

Future research may address some of the limitations in this research experiments. A variety of historical texts from diverse genres, periods, and national languages can be used to increase the validity of the results. Another research direction is to test the developed strategies on texts with real-world OCR mistakes.